# Recycled nylon fibers as cement mortar reinforcement

**Saverio Spadea[a,*], Ilenia Farina[a], Anna Carrafiello[b], Fernando Fraternali[a]**

[a] University of Salerno, Department of Civil Engineering
Via Giovanni Paolo II, 84084 Fisciano (SA), Italy

[b] Omega Plastic srl,
Via Udine 6, 84091 Battipaglia (SA), Italy

[*] Corresponding author: e-mail: sspadea@unisa.it; telephone/fax number: +39 089 964342.

**Abstract**

We investigate engineering applications of recycled nylon fibers obtained from waste fishing nets, focusing our attention on the use of recycled nylon fibers as tensile reinforcement of cementitious mortars. We begin by characterizing the tensile behavior of both unconditioned and alkali-cured recycled nylon fibers obtained through manual cutting of waste fishing net filaments, with the aim of assessing the resistance of such materials to chemical attacks. Special attention is also given to evaluating the workability of fresh mortar and the possible impacts of contaminants released by waste fishing nets into the environment. Next, we deal with compression and bending tests on cementitious mortars reinforced with recycled nylon fibers, and establish comparisons with the experimental behavior of the unreinforced material and with results given in existing literature. In our analysis of different weight fractions and aspect ratios of the reinforcing fibers, we observe marked increases in the tensile strength (up to +35%) and toughness (up to 13 times greater) of the nylon reinforced mortar, as compared with the unreinforced material. The presented results emphasize the high environmental and mechanical potential of recycled nylon fibers for the reinforcement of sustainable cement mortars.

## 1 INTRODUCTION

Protection of the sea environment is one of the most serious issues of this time. In addition to the known causes of environmental degradation, such as pollution, overbuilding of the coast, unconscionable fishing, and coastal erosion, the indiscriminate abandonment of fishing nets on the seabed can cause a growing form of desertification of marine ecosystems. While in the past fishing nets were made of biodegradable natural materials such as cotton and linen, nowadays the nets are typically made of plastic. Fishing net plastics are generally not biodegradable, and therefore it is extremely important to enhance their recycling in order to dispose of wastes and lower the cost of the resulting products. It is worth noting that recent studies have shown that several waste materials can be profitably employed to manufacture low-cost reinforcement techniques of structural and non-structural materials in the construction industry [1]-[4]. In fact, the research in the field of cement mortar is strongly oriented towards the development of suitable materials for the repair and rehabilitation of existing concrete structures [5]-[6].

Polypropylene (PP) and polyamide (PA) fibers have been successfully used in cementitious materials to control shrinkage cracking, to improve material toughness and impact resistance, and to increase significantly the energy absorption capacity of the material [7]-[9]. Habib et al. [10] have carried out an investigation focusing on the effects of synthetic fibers (glass, nylon, and polypropylene) on the mechanical properties of mortars. Such industrial plastic fibers might guarantee better mechanical performance than recycled plastics. However, they inevitably lead to higher energy consumption and emissions. According to the method proposed by the Intergovernmental Panel on Climate Change (IPCC) in 2007 [11], it is estimated that 1.91 kg of equivalent CO2 is needed to produce 1 kg of nylon.

Among recycled plastics, the reinforcement of cementitious materials through recycled polyethylene terephthalate (R-PET) fibers has received particular attention in the technical literature. Several authors have shown that R-PET fibers can conveniently replace virgin plastic fibers in eco-friendly concretes, providing good mechanical and chemical strengths to the final material [12]-[21]. More recently, the R-PET reinforcement of cementitious mortars has also received some attention in the literature [22]-[23]. It should be noted, however, that the use of recycled materials in cementitious mortars remains only very partially investigated. The same holds with respect to the use of nylon fibers as mortar reinforcements. Some recent studies [24]-[25], investigate the recycling of nylon fibers from post-consumer textile carpet waste and their use for concrete reinforcement. In detail, Ogzer and coauthors [25] describe the preparation of nylon fiber-reinforced concrete and the identification of its thermo-mechanical properties, such as compressive and tensile strengths, toughness, specific heat capacity, thermal conductivity, thermal expansion, and hygrometric shrinkage. The results presented in [25] highlight that concretes reinforced with recycled nylon fibers have more ductile and tougher behavior than the unreinforced material, and suffer minor drying shrinkage. Such advantages are, however, balanced by slight reductions of the tensile strength, maximum load-bearing capacity, and modulus of elasticity.

According to the EU green paper [26], waste patches in the Atlantic and the Pacific Oceans are estimated to be in the order of 100 million tons, about 80% of which is plastic. Most of this plastic is due to the indiscriminate abandonment of fishing nets on the seabed, which causes an increasingly common form of desertification of marine ecosystems. A large quantity of fishing nets are also scattered in the Mediterranean Sea or collected on the docks of sea harbors, following seizure by the port authorities.

In the present work, we deal with the reinforcement of a commercial cementitious mortar through recycled nylon (R-nylon) fibers, obtained from waste fishing nets. We begin by assessing that fiber are not potentially harmful for human health by mean of leaching tests. We also perform a preliminary mechanical characterization of the tensile strengths of both unconditioned R-nylon fibers, and alkali-conditioned R-nylon fibers, in order to assess their resistance to chemical attacks. Next, we conduct compression and bending tests on mortar specimens reinforced with R-nylon fibers, comparing the results of such tests to analogous ones referred to the unreinforced mortar. We analyze different fiber weight fractions and aspect ratios of R-nylon fibers. The given results indicate that the examined R-nylon fibers significantly improve the tensile and fracture properties of the base material, as we observe up to 35% increases in tensile strength, and a ductile failure mode in the R-nylon reinforced mortar. The work is completed by comparisons with available literature results on mortars and concretes reinforced through both recycled and virgin plastic fibers.

## 2 MATERIALS AND METHODS

### *2.1 R-nylon fibers and mortar*

We analyzed reinforcing fibers provided by Omega Plastic srl, a company that recovers fishing nets seized by southern Italian port authorities (Anzio, Barletta, Castellabate, Giulianova, Giovinazzo, Lipari, Maratea, Marsala, Marina di Camerota, Milazzo, Mola di Bari, Molfetta, Palermo, Ponza, Sapri, Salina, Trani, and Termoli). After collection, residues of other products are cleaned from the nets. These nets are then classified by type of polymer, cut into pieces, and packed for storage. The final product is generally obtained via a process of extrusion and polymerization performed by plastic material recycling companies. In this case, however, the aim was to analyze a purely mechanical recycling process that does not involve energy consumption and $CO_2$ emissions.

The examined fishing nets are made of aliphatic polyamide 6 (commonly referred to as “nylon 6”). Such a material is widely used in many industrial applications due to its good mechanical and chemical properties, such as, e.g., enhanced toughness and chemical resistance. We performed the manual cutting of the fishing nets in order to obtain R-nylon fibers to be used as mortar reinforcing fibers. At the time of supply, nylon 6 fibers of 0.33 mm diameter were woven into a square mesh with 40 mm sides. We hand-cut this mesh into fibers of the desired length. This entailed cutting 200 mm filaments, each with four knots, to be used for uniaxial characterization tests. We also hand-cut short fibers of different lengths (12.7 mm, 25.4 mm and 38.2 mm), to be employed in the mortar-reinforcing process (See Table 2 for details).

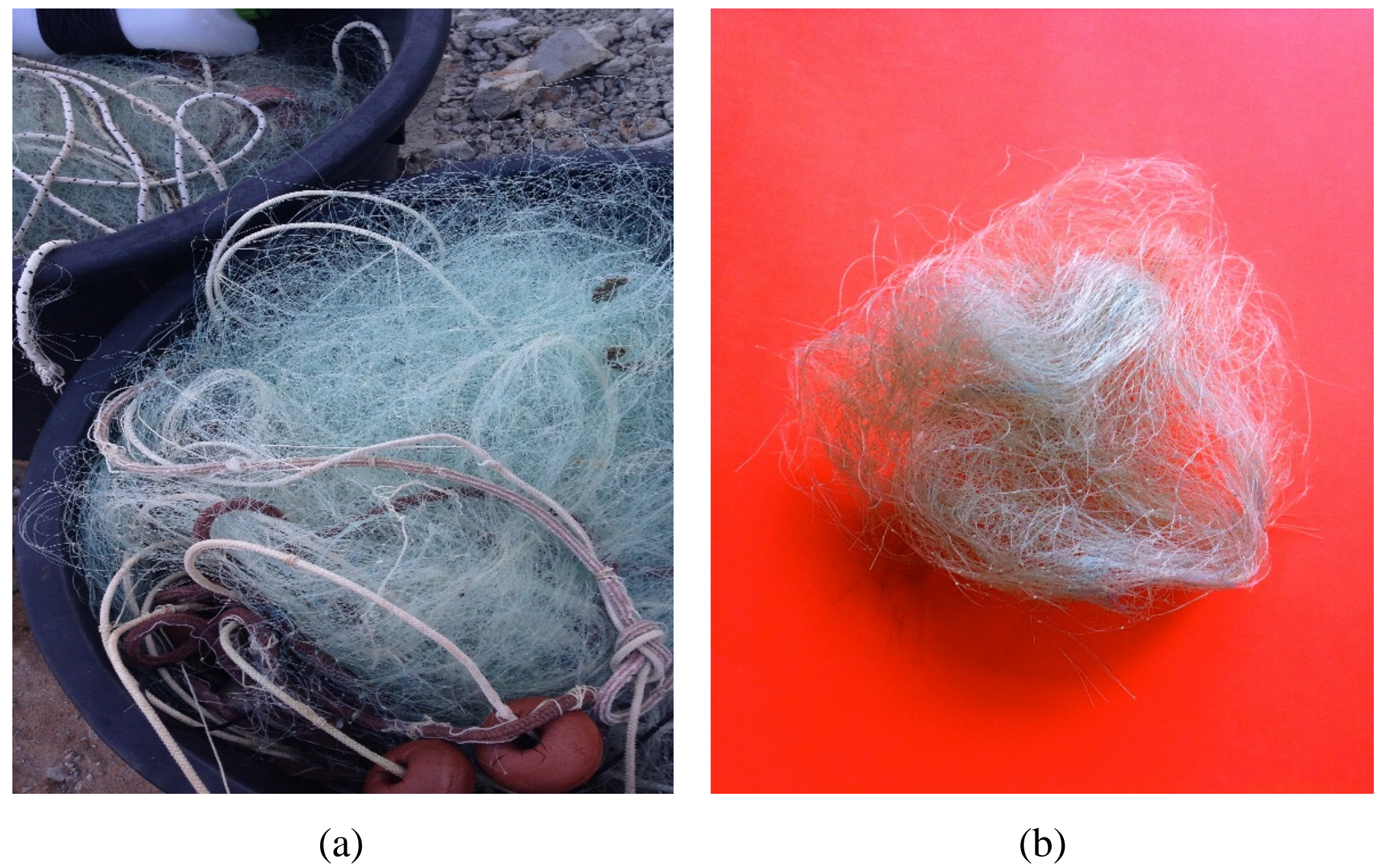

(a) (b)

**Figure 1 – Examined nylon 6 waste fishing nets.**

We employed the commercial mortar *Disbocret Unitech R4*, produced by Italian Caparol GmbH & Co, for the sake of comparing the present results with those deriving from the reinforcement of the same kind of mortar with PET stripes [18]. This product is aimed at repairing damaged concrete, and includes PVC micrometric reinforcing fibers (diameter of about 50 μm, length of about 1 mm) aimed at enhancing material thixotropy and shrinkage resistance rather than mechanical properties. According to the producer datasheet, the mortar owes the following mechanical properties:

- strength class R4, e.g. compressive strength greater than 45 MPa (EN 1504);
- bond to existing concrete greater than 2 MPa (EN 1542);
- Young's modulus of elasticity greater than 20 GPa (EN 13412).

### *2.2 Leaching tests on waste fishing nets*

The leaching test is intended to simulate the release of contaminants by placing a reagent in contact with a leaching agent for a defined period of time. In the present case, we conducted the test on 100 gr of the disposed fishing net (cut into small pieces with a maximum length equal to 4 mm), without previously washing the fibers. As per the standard leaching test EN 12457-2 [27], they were placed into an agitation apparatus consisting of one liter of $CO_2$-saturated water for 24 hours at a temperature of 20 ± 5 °C. Finally, the liquid was filtered in order to obtain the eluate, which we subjected to chemical analysis. The results of the test are shown in Table 1, together with the limit values in UNI 10802:2013 [28] – the standard giving the acceptance criteria for recycled aggregates in construction. These results show that R-Nylon fibers may not lead to overall adverse environmental or human health impacts. As such, they can be safely used in the reinforcing phase for cement materials.

**Table 1 – Results of Leaching Tests on R-nylon fibers**

| Compound | | current value | limit value |
|---|---|---|---|
| Nitrati | mg/l | 5.9 | 50 |
| Fluoruri | mg/l | <0,1 | 1.5 |
| Solfati | mg/l | 3.6 | 250 |
| Cloruri | mg/l | 2.2 | 100 |
| Cianuri | mg/l | <0,1 | 50 |
| Bario | mg/l | <0,1 | 1 |
| Rame | mg/l | <0,01 | 0.05 |
| Zinco | mg/l | <0,1 | 3 |
| Berillio | µg/l | <0,1 | 10 |
| Cobalto | µg/l | <0,1 | 250 |
| Nichel | µg/l | 0.3 | 10 |
| Vanadio | µg/l | <0,1 | 250 |
| Arsenico | µg/l | <0,1 | 50 |
| Cadmio | µg/l | <0,1 | 5 |
| Cromo | µg/l | 0.3 | 50 |
| Piombo | µg/l | 0.2 | 50 |
| Selenio | µg/l | <0,1 | 10 |
| Mercurio | µg/l | <0,1 | 1 |
| Amianto | mg/l | <0,01 | 30 |
| COD | mg/l | 12.6 | 30 |
| PH | - | 8.2 | >5,5 <12,0 |

### *2.3 Alkali conditioning of R-nylon fibers*

We conditioned 200 mm R-nylon monofilaments in an alkaline environment according to the ASTM D543-06 standard [27]. In detail, the filaments were cured in a solution consisting of 10.4 g of sodium hydroxide immersed in 999 ml of distilled water for 120 hours (5 days), keeping the temperature constant at 60 ± 2 °C through a climatic chamber.

We weighted the fibers using an electronic scale (resolution 1 mg) and inspected them with an optical microscope (ZEISS Axioskop 40) using 5× magnification both before and after the treatment. We observed a slight loss of mass, equal to about 1.7%, but no relevant signs of corrosion on the fiber surface (cf. Figure 2). By zooming in the images acquired from the optical microscope, we were able to consistently measure the geometry of the fibers. As shown in Figure 2, the cross-section appears circular with 0.33 mm diameter, for both conditioned and unconditioned R-nylon fibers.

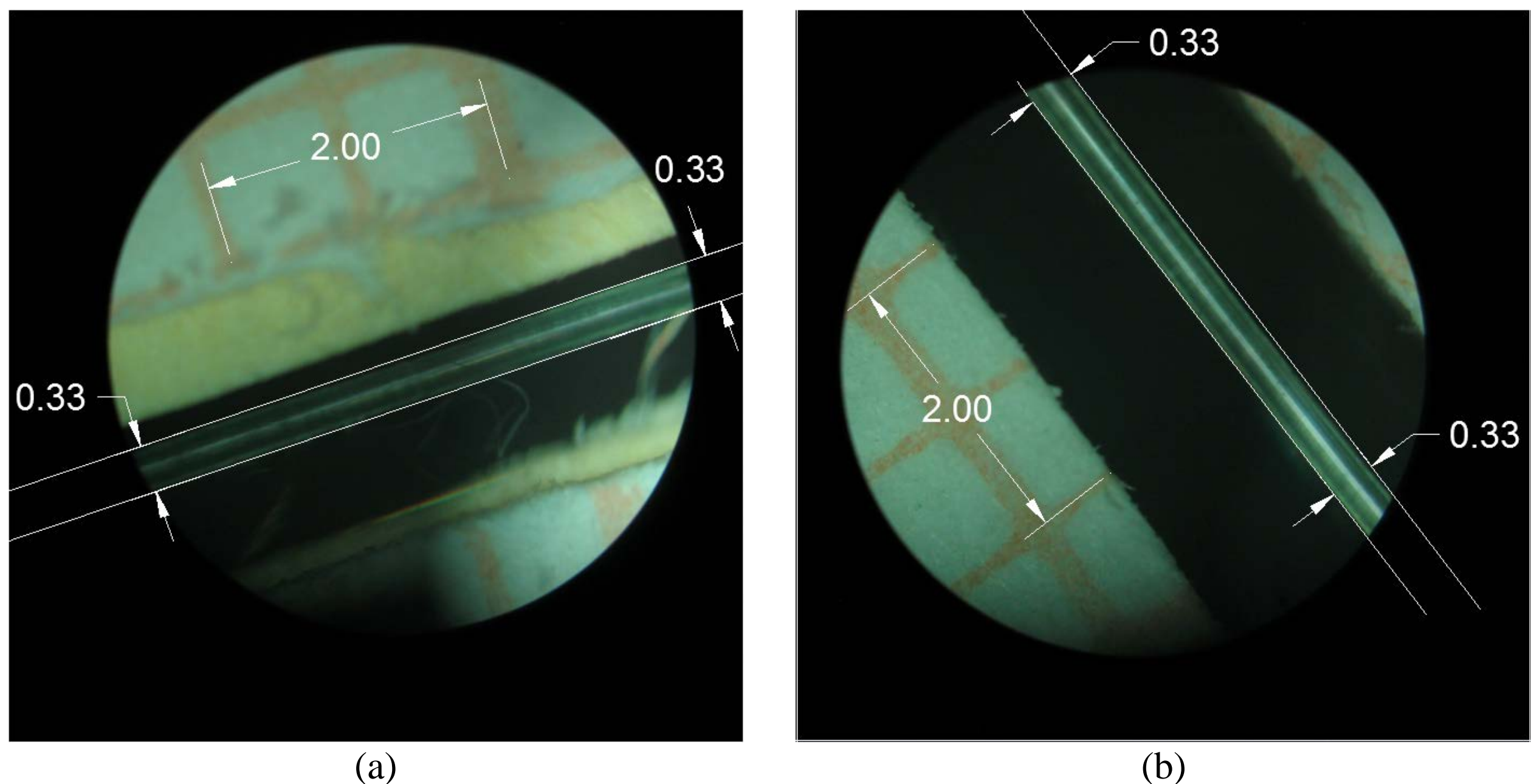

(a) (b)

**Figure 2 – Optical microscope images (length units: mm): a) alkali-conditioned R-nylon fibers; unconditioned R-nylon fibers.**

## *2.4 Preparation of Mortar Samples*

The preparation of mortar prismatic samples included the following steps:

- weighting of mortar and nylon fibers;
- hand mixing of dry materials in order to uniformly distribute fibers into the mortar premix;
- hydrating the mixture by adding the target quantity of water (180 g of water for each kilogram of mortar);
- mechanical shaking of the mixture at slow speed (for about two minutes) until a homogeneous and workable product of semi-fluid consistency was obtained;
- casting the prismatic specimens into 40 mm × 40 mm × 160 mm molds, accurately vibrating them and covering them with a plastic sheet.

We removed the specimens from the molds after 24 h curing at room temperature. Next, we cured the unmolded specimens in water at 23 °C for 28 days until testing. We examined two plain mortar specimens, and two specimens in correspondence with six different fiber-reinforcement mixes, thus obtaining a total of 14 prismatic specimens. As described in Table 2, the examined mortar mixes vary due to the content and aspect ratio of R-nylon fibers. Throughout the paper, we refer to the plain mortar specimens as “UR” and the reinforced mortar specimen as “PA – fiber length in inches – weight fraction” (Table 2).

The workability of fresh mortar mixes was determined by flow table test as per EN 1015-3 [34]. As described in Table 1, the content and aspect ratio of R-nylon fibers influences the properties of fresh mortar mixes. The greater the amount of fibers, and the greater their length, the lower the workability of the fresh mix.

**Table 2 – Specimens designation, mixes and flow table test**

| Specimens | | | Fibers | | | Flow |
|---|---|---|---|---|---|---|
| Designation | Quantity | Fiber fraction | Length *l* | Diameter *d* | Aspect ratio (*L*/*d*) | Diameter D |
| | | % in weight | mm (inches) | mm | – | mm |
| UR | 2 | – | – | – | – | 176 |
| PA-0.5-1.0% | 2 | 1.0 | 12.7 (0.5) | 0.33 | 38.5 | 165 |
| PA-0.5-1.5% | 2 | 1.5 | 12.7 (0.5) | 0.33 | 38.5 | 155 |
| PA-1.0-1.0% | 2 | 1.0 | 25.4 (1.0) | 0.33 | 77.0 | 150 |
| PA-1.0-1.5% | 2 | 1.5 | 25.4 (1.0) | 0.33 | 77.0 | 161 |
| PA-1.5-1.0% | 2 | 1.0 | 38.1 (1.5) | 0.33 | 115.5 | 143 |
| PA-1.5-1.5% | 2 | 1.5 | 38.1 (1.5) | 0.33 | 115.5 | 138 |

## 3 MECHANICAL TESTS AND RESULTS

### *3.1 Uniaxial tensile tests on R-nylon fibers*

We carried out uniaxial tensile tests according to ASTM C1557-03 Standard [30] on both conditioned and unconditioned 200 mm R-nylon filaments in order to determine the tensile strength and Young's modulus of such elements. The tests were performed under a constant cross-head displacement rate by means of an MTS SANS testing machine equipped with a 1 kN load cell and pneumatic grips. We set testing rates that were sufficiently rapid to possibly obtain the failure strength within 30 s.

With the aim of accurately defining the gage length and preventing eccentricity of the load and fiber twisting, we preliminarily mounted the specimens on cardboard tabs and cut the tabs sides after fastening the specimens. As shown in Figure 3, three different gage lengths $L_0$ were adopted (1 in, 1¼ in, and 1½ in). During the tests we continuously acquired force and cross-head displacement, observing a linear elastic behavior of the tested specimens up to failure.

The peak force $F_{max}$ and the elongation to force ratios $\Delta L / F$ exhibited by conditioned and unconditioned fibers are shown in Table 3 and Table 4, respectively. The quantity $\Delta L / F$ was derived using a linear regression for the Force–Displacement segment between 20 and 50% of the failure load. The mean, standard deviation (*SD*), and variance (*CV*) of the tensile stresses are also given in Tables 2 and 3 for each group of specimens.

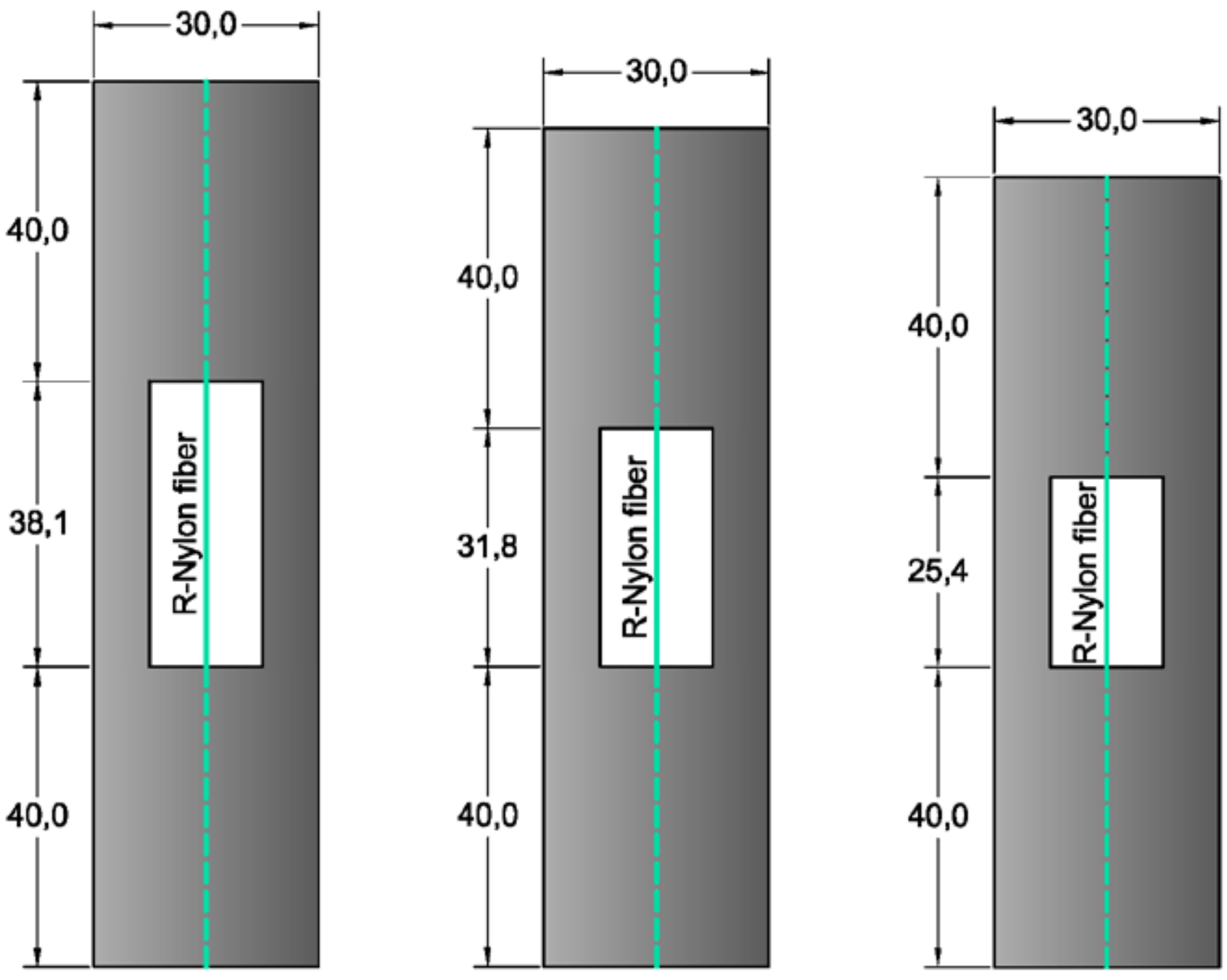

**Figure 3 – Cardboard mounting tabs (units: mm)**

**Table 3 – Results of tensile tests on unconditioned mortar specimens**

| **Gage length** $L_0$ | | **1** | | **2** | | **3** | | **Tensile strength** | | |
|---|---|---|---|---|---|---|---|---|---|---|
| | | $F_{max}$ | $\Delta L/F$ | $F_{max}$ | $\Delta L/F$ | $F_{max}$ | $\Delta L/F$ | **Mean** | *SD* | *CV* |
| Inches | mm | kN | mm/kN | kN | mm/kN | kN | mm/kN | MPa | MPa | % |
| 1 | 25.4 | 27.0 | 0.32 | 30.6 | 0.33 | 29.2 | 0.31 | 338 | 21 | 6.3% |
| 1¼ | 31.8 | 30.5 | 0.38 | 31.6 | 0.41 | 27.2 | 0.47 | 348 | 27 | 7.7% |
| 1½ | 38.1 | 25.4 | 0.51 | 27.2 | 0.49 | 24.8 | 0.53 | 302 | 15 | 4.8% |

**Table 4 – Results of tensile tests on conditioned mortar specimens**

| **Gage length** $L_0$ | | **1** | | **2** | | **3** | | **Tensile strength** | | |
|---|---|---|---|---|---|---|---|---|---|---|
| | | $F_{max}$ | $\Delta L/F$ | $F_{max}$ | $\Delta L/F$ | $F_{max}$ | $\Delta L/F$ | **Mean** | *SD* | *CV* |
| Inches | mm | kN | mm/kN | kN | mm/kN | kN | mm/kN | MPa | MPa | % |
| 1 | 25.4 | 24.8 | 0.30 | 28.5 | 0.31 | 28.6 | 0.32 | 319 | 25 | 7.9% |
| 1¼ | 31.8 | 24.8 | 0.37 | 30.1 | 0.39 | 32.8 | 0.40 | 342 | 48 | 13.9% |
| 1½ | 38.1 | 21.7 | 0.50 | 26.7 | 0.53 | 25.8 | 0.49 | 289 | 31 | 10.8% |

In the case of 1 in and 1¼ in specimens, we observed satisfactory failure modes and very low dispersion in peak loads. The mean values of the tensile strength results were 338 and 348 MPa respectively for unconditioned specimens and 319 and 342 MPa for conditioned ones. In contrast, the 1½ in specimens showed a considerably lower tensile strength and failure of fibers in the vicinity of grips. This was most probably due to the proximity of the knots to the extreme points of the gage length, which cause local decrease of the fiber mechanical properties.

By restricting our analysis on 1 in and 1¼ in specimens (i.e., neglecting the 1½ in specimens), we observed a very slight decrease in the average tensile strength (about –4%) in conditioned specimens (343 MPa) with respect to unconditioned ones (330 MPa).

We completed the fiber characterization by determining the Young modulus $E$ of such elements through the procedure suggested by ASTM C1557-03 [30]. The adopted method consists of plotting the $\Delta L / F$ experimental values, which provide the inverse of the slope of the force versus cross-head displacement curve, against the corresponding geometric parameter $L_0 / A$, with $A$ denoting the cross-section area. Assuming a linear elastic constitutive law and uniaxial load conditions, the following analytical expression applies:

$$\frac{\Delta L}{F} = \frac{1}{E}\frac{L_0}{A} + C_S \,. \tag{1}$$

According to Eqn. (1), the linear regression of the experimental data $\Delta L / F$ vs. $L_0 / A$ (cf. Figure 4) yields a straight line with a constant slope $1/E$ (the inverse of Young modulus) and the intercept $C_S$ giving the system compliance. The above procedure allowed us to determine the average experimental values of the Young moduli, which were found equal to 728 MPa for unconditioned fibers and 724 MPa for the conditioned fibers. The observed minor decreases of tensile strengths and Young moduli due to the alkali conditioning lead us to conclude that the examined R-nylon fibers show excellent alkali resistance. There is some evidence in the literature that absorbed water may considerably affect the mechanical behavior of nylon 6, causing stiffness drops together with toughness improvement [31]-[32]. Such a phenomenon is demonstrated in the present case by the low Young modulus of the R-nylon fibers (724-728 MPa), considering that nylon 6 at the virgin state typically exhibits Young modulus in the range 1-3 GPa [33].

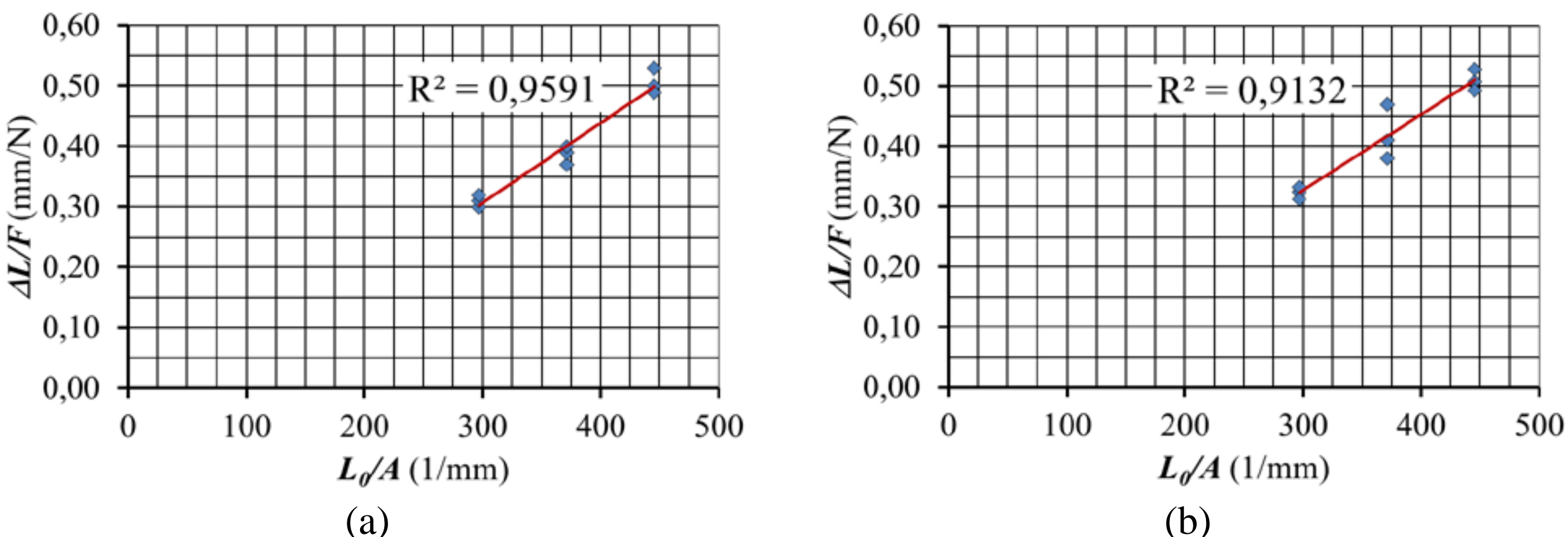

**Figure 4 – $\Delta L / F$ vs $L_0 / A$ plots: a) Unconditioned specimens; b) Conditioned specimens.**

### *3.2 Bending and Compressive Tests*

We performed three-point load-bending tests and compression tests on mortar prismatic specimens according to EN 1015-11:2007 [35] using a MATEST electrically operated testing machine with a 200 kN capacity (Figure 5). Bending tests were carried out under displacement control at a speed of 0.01 mm per second on a clear span of 100 mm. Compression tests were performed on the two portions resulting from the failure the specimens subjected to bending tests. We ended up with compression tests on ideal cubes with dimensions of 40 mm × 40 mm × 40 mm, applying a loading rate equal to 100 N per second.

As expected, the unreinforced mortar specimen exhibited brittle behavior in bending due to a sudden rupture on reaching the peak load. On the other hand, fiber-reinforced mortar specimens showed considerable post-peak resources.

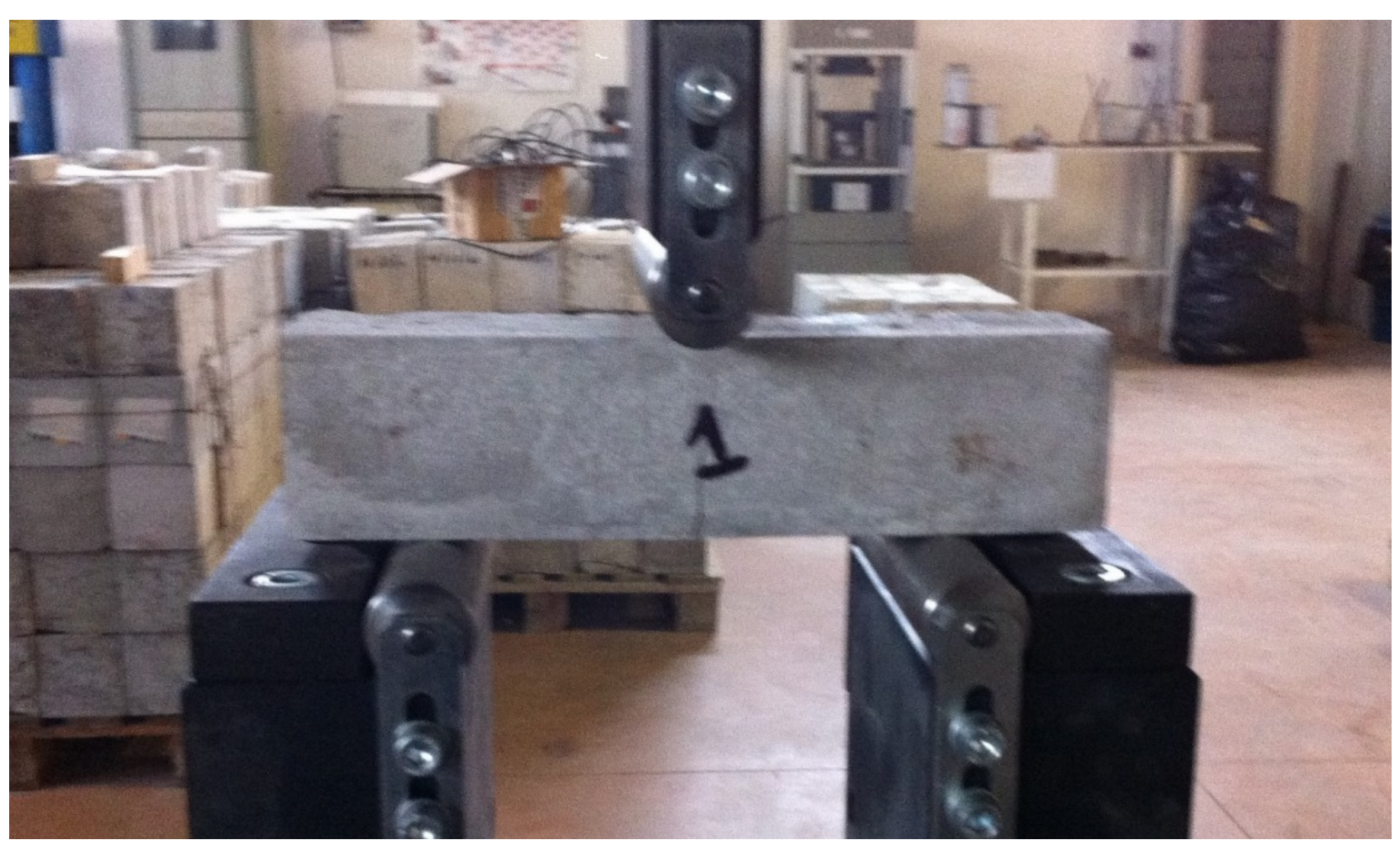

**Figure 5 – Three-point bending test setup.**

Load–deflection curves ( $P-\delta$ ) obtained during the bending test are shown in Figures 6, 7, and 8, with reference to the specimen reinforced by 0.5, 1.0, and 1.5 in fibers, respectively. In the same chart we report the curves referring to mixtures containing different percentages of R-nylon fibers (1.0 and 1.5% by weight).

Typically, the tests show a drop of the load after reaching the peak load. Starting from the post-peak load target, the curves exhibit an almost constant branching or a hardening kind of constitutive behavior depending on the length and quantity of fibers. More specifically, we observe the following:

- a higher percentage of fibers (1.5 rather than 1.0%) causes a less relevant drop of load after the peak value (see Figures 6 and 8);
- fibers of greater length (1.0 and 1.5 in rather than 0.5 in) confer a hardening type of post-peak behavior on the mortar.
- while the peak loads are quite consistent for specimens within the same mix, some variation in post-peak behavior can be observed (see Figures 6 and 8). An influential factor here may be the section in which the fracture initiates. It is not always the case that the fracture occurs right in the midspan.

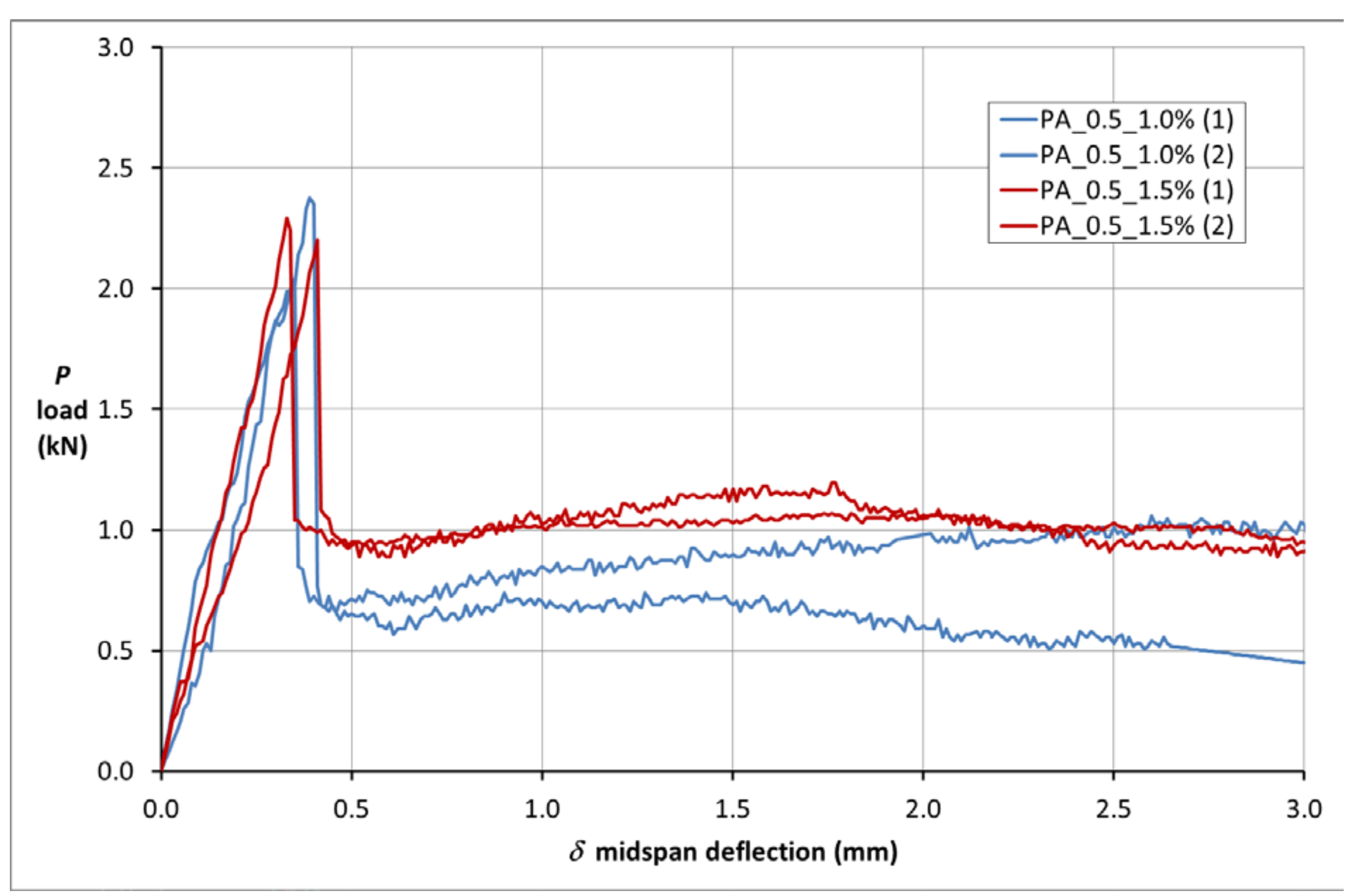

**Figure 6 – Load–deflection curves of mortar specimens reinforced with 0.5 in R-nylon fibers.**

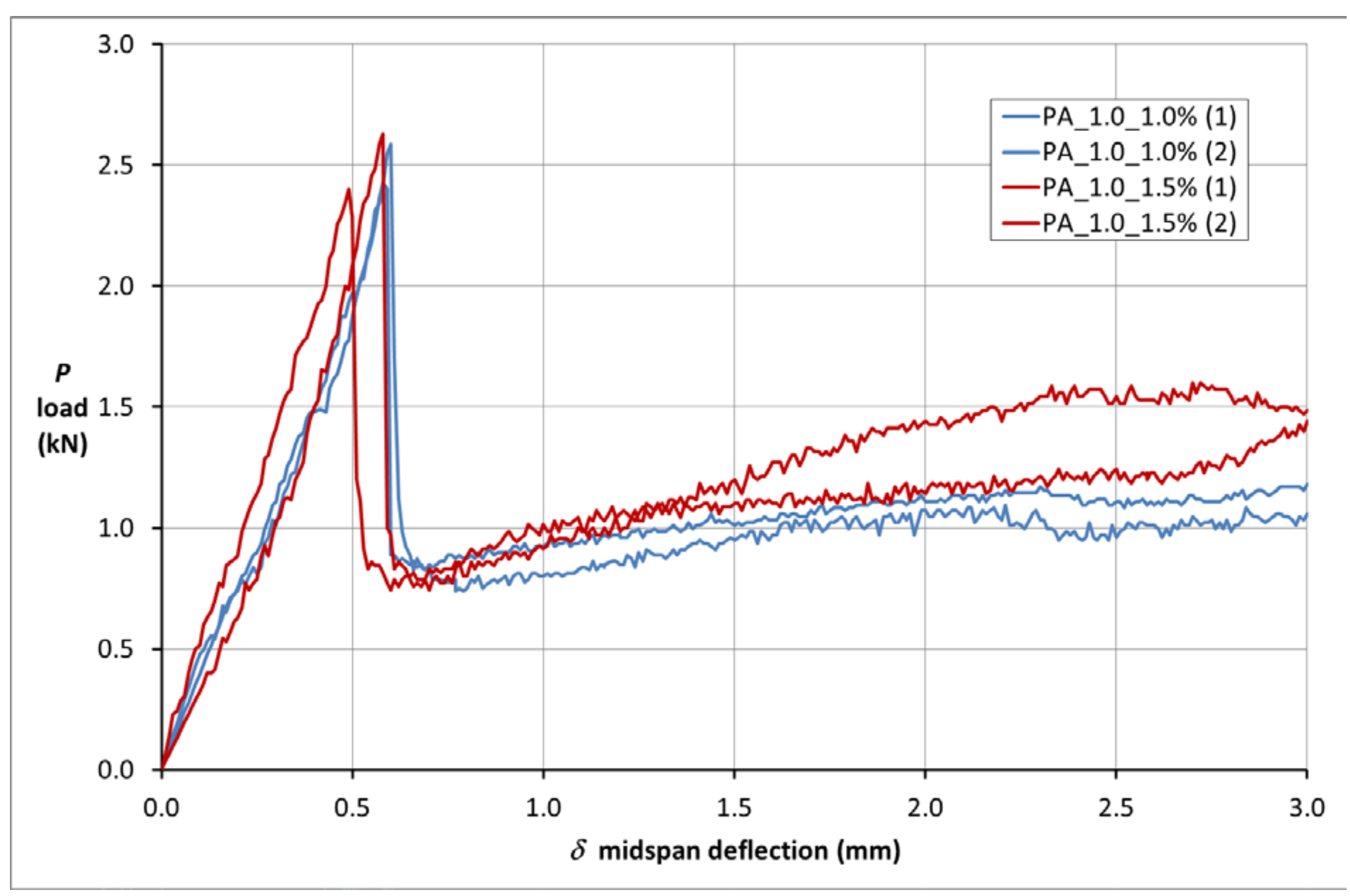

**Figure 7 – Load-deflection curves of mortar specimens reinforced with 1.0 in R-nylon fibers.**

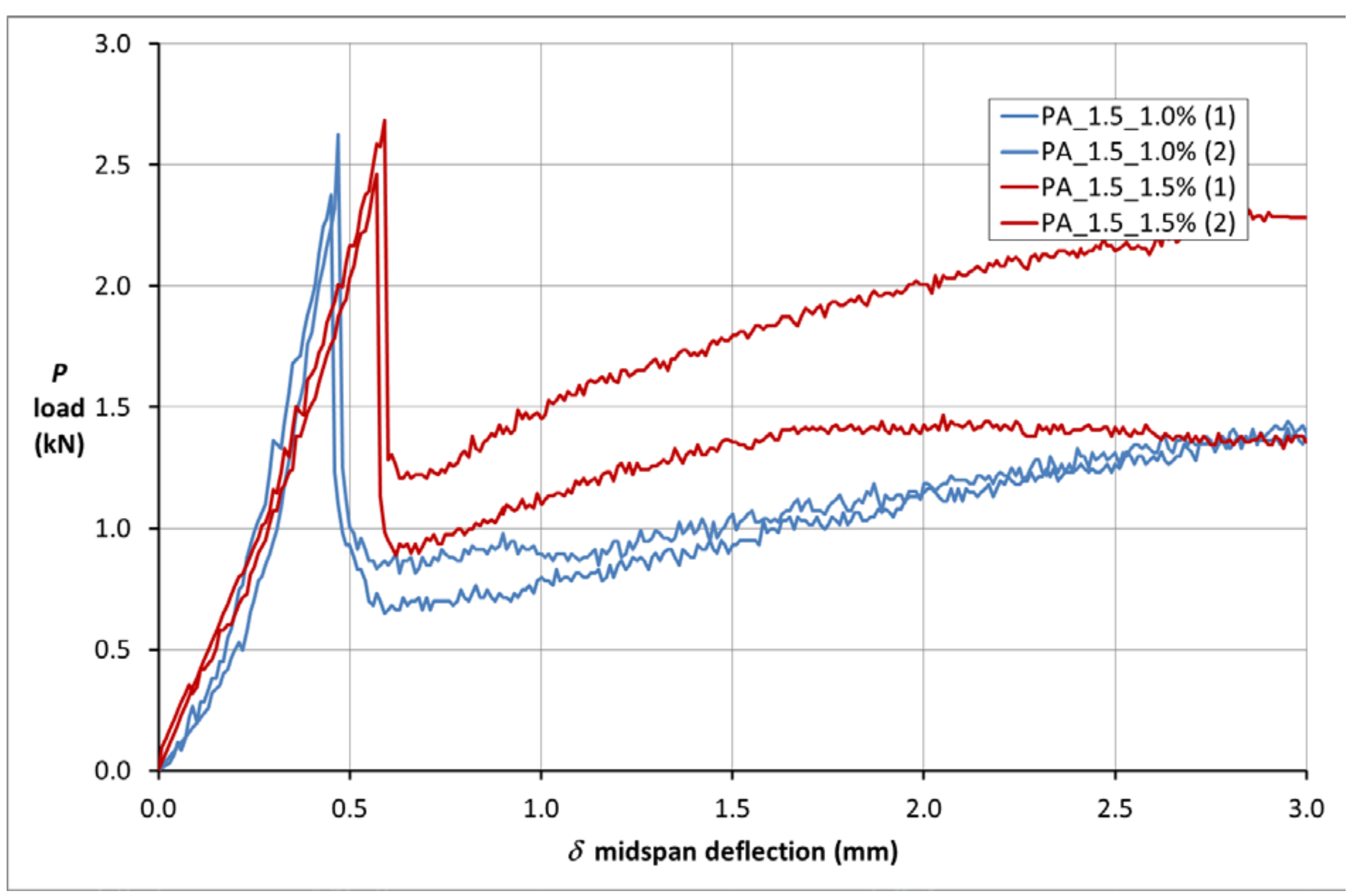

**Figure 8 – Load-deflection curves of mortar specimen reinforced with 1.5 in R-nylon fibers.**

Table 5 shows the peak load ($P_{cr}$) exhibited by each specimen, the corresponding midspan deflection ($\delta_{cr}$), and the values of first crack strength, $f_{cr}$, calculated as follows (EN 1015-11:2007):

$$f_{cr} = \frac{3}{2}\frac{Pl}{a^3}. \quad (2)$$

For each examined mixture, we also report the mean, standard deviation (*SD*), and coefficient of variance (*CV*) of the first crack strength. In the last column of Table 4 we show the percentage change in the first crack strength in reinforced mortars with respect to the unreinforced ones ($\Delta f_{cr}$).

Table 5 and Figure 9 highlight how reinforcing fibers confer a considerably higher resistance to cracking (up to +35%). Furthermore, it is evidenced that longer fibers are most effective, as the increase in strength varies from 16–18% to 32–35% when fibers of 1.0 and 1.5 in are used instead of those measuring 0.5 in. Contrarily, the weight ratio of fibers does not seem to be an important factor in regard to this. Such results bear special relevance, considering that reductions of the first crack strength over the unreinforced material were instead observed in the case of a R-PET reinforcement of the same mortar analyzed in the present study (up to –20% reductions of $f_{cr}$, [23]), and for the R-nylon reinforcement of a concrete based on a Portland limestone cement and a 0.35 water/cement ratio (about -11% first-crack strength reductions in four-point bending tests, [25]).

**Table 5 – Results of three-point bending tests**

| *Specimen* | $P_{cr}$ kN | $\delta_{cr}$ mm | $f_{cr}$ MPa | *mean* MPa | *SD* MPa | *CV* % | $\Delta f_{cr}$ % |
|---|---|---|---|---|---|---|---|
| UR (1)<br>UR (2) | 2.00<br>1.80 | 0.32<br>0.39 | 3.61<br>4.23 | 4.46 | 0.33 | 7 | – |
| PA-0.5-1.0% (1)<br>PA-0.5-1.0% (2) | 2.38<br>2.04 | 0.39<br>0.35 | 5.57<br>4.79 | 5.18 | 0.55 | 11 | +16 |
| PA-0.5-1.5% (1)<br>PA-0.5-1.5% (2) | 2.29<br>2.20 | 0.33<br>0.41 | 5.37<br>5.17 | 5.27 | 0.15 | 3 | +18 |
| PA-1.0-1.0% (1)<br>PA-1.0-1.0% (2) | 2.43<br>2.59 | 0.58<br>0.60 | 5.69<br>6.06 | 5.87 | 0.27 | 5 | +32 |
| PA-1.0-1.5% (1)<br>PA-1.0-1.5% (2) | 2.40<br>2.63 | 0.49<br>0.58 | 5.63<br>6.16 | 5.89 | 0.38 | 6 | +32 |
| PA-1.0-1.5% (1)<br>PA-1.0-1.5% (2) | 2.38<br>2.63 | 0.49<br>0.47 | 5.57<br>6.15 | 5.86 | 0.41 | 7 | +32 |
| PA-1.0-1.5% (1)<br>PA-1.0-1.5% (2) | 2.69<br>2.47 | 0.59<br>0.56 | 6.29<br>5.77 | 6.03 | 0.37 | 6 | +35 |

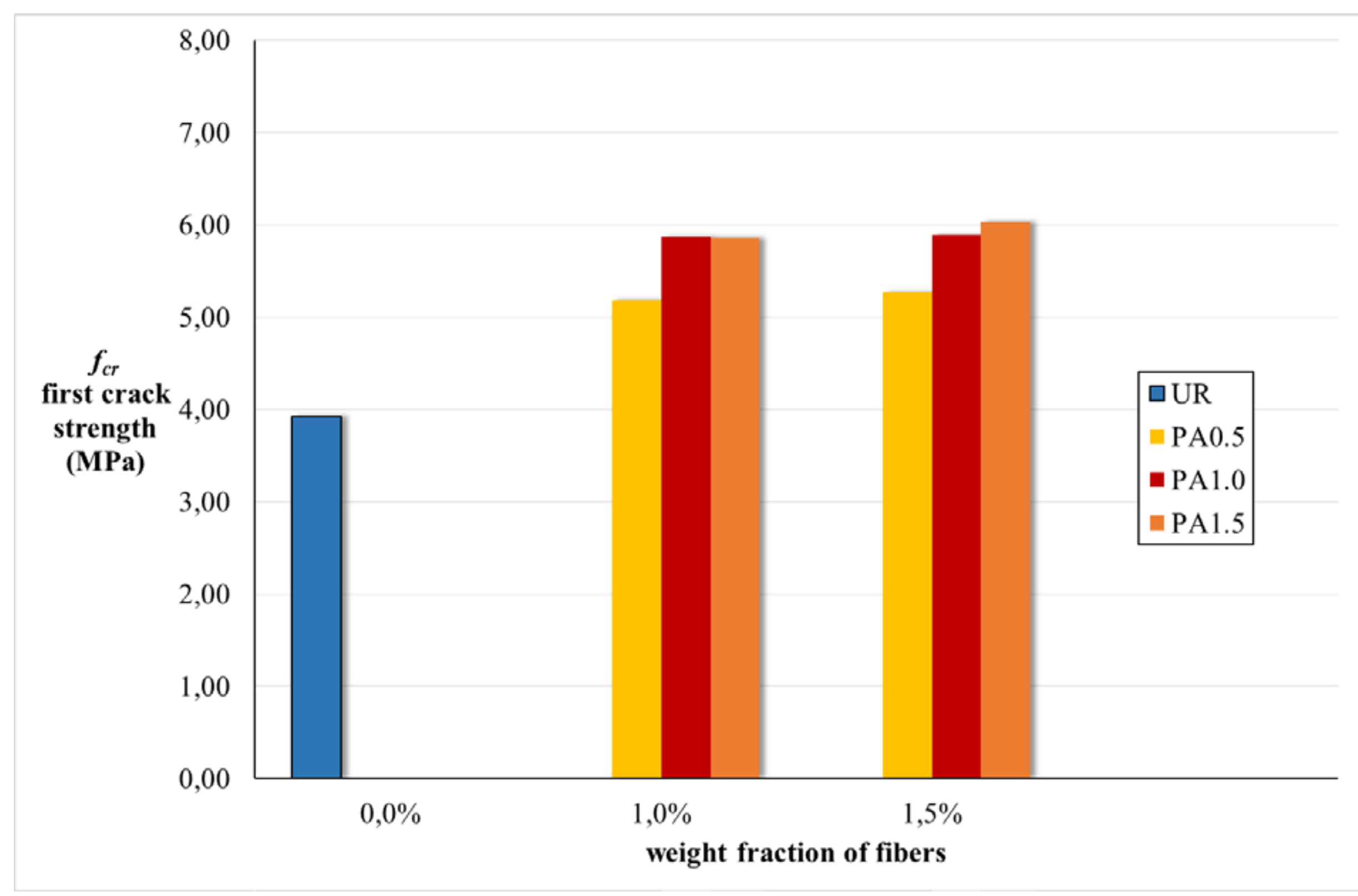

**Figure 9 – Average first crack strengths of the analyzed mortars.**

With reference to virgin fibers, Habib et al [10] observed greater improvements in the first crack strength (up to +149%) in reinforced mortar specimens as compared to R-Nylon. As shown in Table 6, incremental differences in strength are observed, especially when increasing the fiber content from 0.5% to 1.0%.

**Table 6 – Results of bending tests performed by Habib et al [10]**

| **fiber content** (%) | 0.5 | 1.0 | 2.0 | 0.5 | 1.0 | 2.0 | 0.5 | 1.0 | 2.0 |
|---|---|---|---|---|---|---|---|---|---|
| **fiber length** (in) | | 0.5 | | | 1.0 | | | 1.5 | |
| $\Delta f_{cr}$ (%) | +57 | +112 | +120 | +87 | +95 | +149 | +14 | +108 | +112 |

Table 7 shows the cube compressive strength, $f_c$, calculated as required by EN 1015-11:2007 [35]. For each kind of mixture considered, we also report the mean, standard deviation (*SD*), and variance (*CV*) of the compressive strength. In the last column of Table 7 we show the percentage change of compressive strength in reinforced mortars with respect to unreinforced ones ($\Delta f_c$). The results in Table 7 and Figure 10 show that the addition of R-nylon fibers causes a decrease in the compressive strength of the examined mortar (up to –37%), especially when very short fibers are employed.

It is worth noting that the same specimens were used to perform both bending and compressive tests. This practices does not affect the results in terms of compressive strength for ordinary mortar specimens because they fail in bending as soon as the first crack occurs. However, the heavily cracked regime experienced by reinforced mortar specimens during bending tests may compromise the integrity of their microstructure, affecting the results of compressive tests.

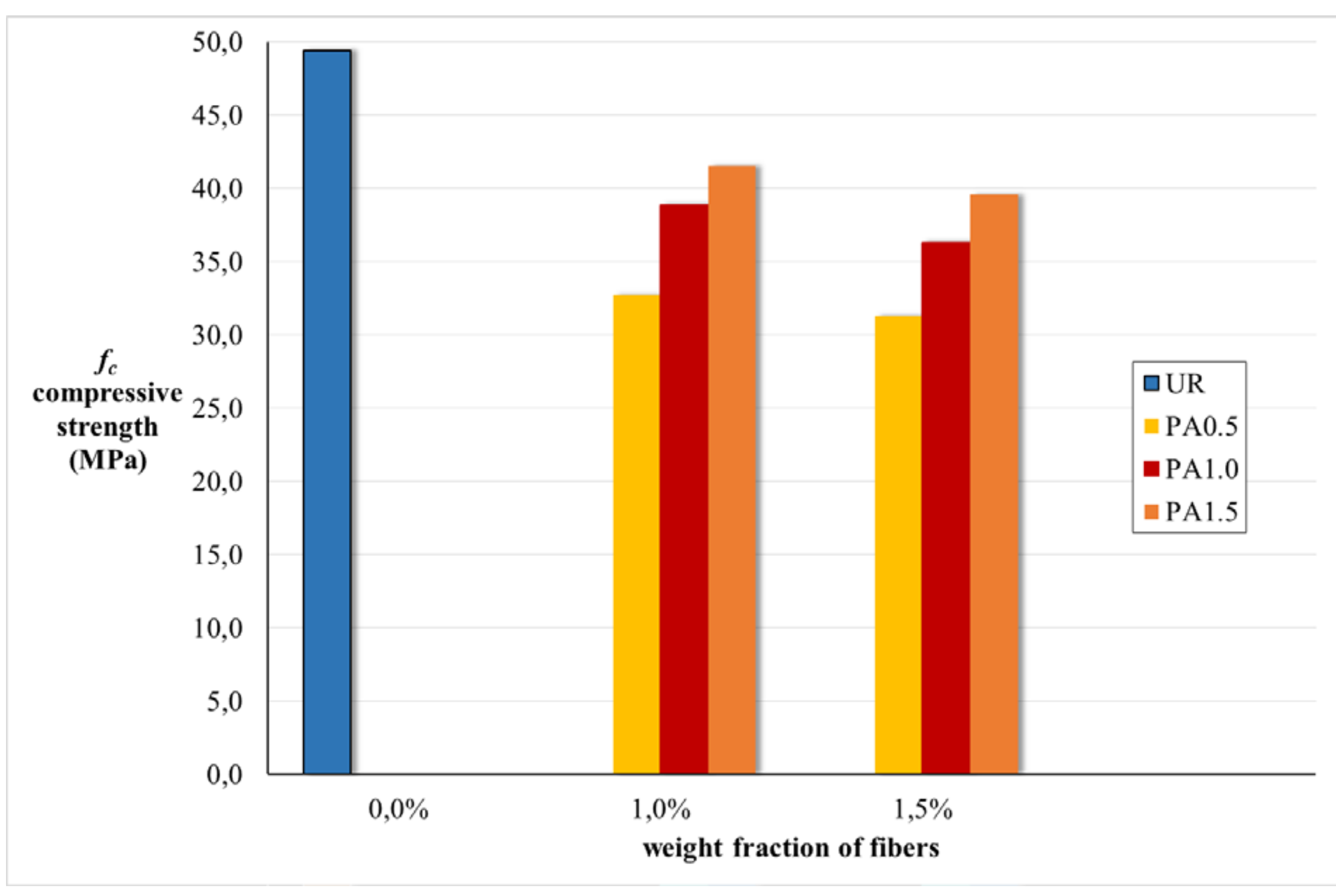

**Figure 10 – Average compressive strengths of the analyzed mortars.**

**Table 7 – Results of compression tests**

| *Specimen* | $f_c$<br>MPa | *Mean*<br>N | *SD*<br>N | *CV*<br>% | $\Delta f_c$<br>% |
|---|---|---|---|---|---|
| UR (1) | 43.9 | 49.4 | 3.7 | 7 | |
| | 50.4 | | | | |
| UR (2) | 51.6 | | | | |
| | 51.6 | | | | |
| PA-0.5-1.0% (1) | 34.6 | 32.7 | 2.0 | 6 | –34 |
| | 29.9 | | | | |
| PA-0.5-1.0% (2) | 33.8 | | | | |
| | 32.5 | | | | |
| PA-0.5-1.5% (1) | 32.7 | 31.3 | 1.0 | 3 | –37 |
| | 31.3 | | | | |
| PA-0.5-1.5% (2) | 31.1 | | | | |
| | 30.2 | | | | |
| PA-1.0-1.0% (1) | 40.9 | 38.9 | 2.0 | 5 | –21 |
| | 37.4 | | | | |
| PA-1.0-1.0% (2) | 40.2 | | | | |
| | 37.0 | | | | |
| PA-1.0-1.5% (1) | 38.3 | 36.3 | 2.4 | 7 | –27 |
| | 32.8 | | | | |
| PA-1.0-1.5% (2) | 36.8 | | | | |
| | 37.3 | | | | |
| PA-1.0-1.5% (1) | 39.9 | 41.5 | 3.2 | 8 | –16 |
| | 43.6 | | | | |
| PA-1.0-1.5% (2) | 44.8 | | | | |
| | 37.9 | | | | |
| PA-1.0-1.5% (1) | 38.6 | 39.6 | 1.9 | 5 | –20 |
| | 42.4 | | | | |
| PA-1.0-1.5% (2) | 39.3 | | | | |
| | 38.1 | | | | |

The effect of reinforcing plastic fibers in compression is quite often debated in the literature, since different studies report increases [3],[7],[8],[10],[15], decreases [2],[9],[14],[18] or no relevant effects [16],[21],[25] in the compressive strength of fiber-reinforced concretes and mortars, with respect to the unreinforced materials. The decrease of compressive properties has been explained due to the fact that highly deformable plastic fibers might assume the role of voids in the cementitious matrix when compressive forces are applied [9],[14],[18]. On the other hand, it has been shown that fibers of sufficient stiffness and length can increase the lateral tensile strength of the composite material, leading to delayed compression failure [7],[8],[10],[25].

In particular, Habib et al [10] observed marked increases of the compressive strength (up to +75%) in mortar specimens reinforced with industrial nylon fibers with respect to their unreinforced counterparts. As shown in Table 8, incremental differences in strength are observed both when increasing the fiber content from 0.5% to 2.0% and when increasing the fiber length from 0.5 in to 1.5 in.

As we already observed, the present R-nylon fibers suffered a noticeable stiffness drop due to water absorption (cf. Sect. 3.1), and indeed we observe decreases in the compressive strengths of the fiber-reinforced mortars, as compared with the unreinforced material. Such strength drops are nevertheless less relevant in the case of longer fibers, due to a major anchor length (cf. Figure 10).

**Table 8 – Results of compression tests performed by Habib et al [10]**

| **fiber content** (%) | 0.5 | 1.0 | 2.0 | 0.5 | 1.0 | 2.0 | 0.5 | 1.0 | 2.0 |
|---|---|---|---|---|---|---|---|---|---|
| **fiber length** (in) | | 0.5 | | | 1.0 | | | 1.5 | |
| $\Delta f_c$ (%) | +25 | +28 | +58 | +43 | +40 | +58 | +40 | +58 | +75 |

### *3.3 Toughness indices and residual strength factors*

The toughness exhibited by mortar specimens in three-point bending tests is typically measured by computing the area under the corresponding load–deflection curve, which is related to the energy absorption capacity of the material. The following equation denotes the area under the load-deflection curve up to a target value of deflection $\bar{\delta}$:

$$T_{\bar{\delta}} = \int_0^{\bar{\delta}} P\delta \, d\delta \, , \tag{3}$$

Although ASTM C 1609-06 [36] is currently accepted as the test method for the analysis of fiber-reinforced concrete, we hereafter consider the ASTM C 1018-97 [37] procedure to analyze the load deflection curve of mortar specimens. Both standards refer to a three-point bending test but the procedure required by ASTM C 1018-97 [37] is better suited to small size specimens, such as the mortar specimens examined in the present study. In line with such a standard, we introduce the following toughness indices:

$$I_5 = \frac{T_{3\delta_{cr}}}{T_{\delta_{cr}}} \, , \tag{4}$$

$$I_{10} = \frac{T_{5.5\delta_{cr}}}{T_{\delta_{cr}}} \, , \tag{5}$$

where $\delta_{cr}$ is the deflection corresponding to the first crack strength (Table **5**).

We also introduce the residual strength factor:

$$R_{a,b} = \frac{100}{(a-b)} \left( I_a - I_b \right) , \tag{6}$$

which is related to the load-carrying capacity of the material after crack onset. Assuming a linear force–deflection response up to $\delta_{cr}$ it can be easily verified that a perfectly plastic post-crack behavior corresponds to $R_{a,b} = 100$.

Table 9 shows the flexural toughness exhibited* by each specimen at three different values of deflection ($\delta_{cr}$, $3\delta_{cr}$, $5.5\,\delta_{cr}$). We also report the average values of toughness indices $I_5$ and $I_{10}$ as well as the residual strength factor:

$$R_{10,5} = 20\left(I_{10} - I_5\right). \tag{7}$$

Toughness indices and residual strength factors could not be computed for the UR mortar, due to the brittle behavior of such material.

**Table 9 – Flexural toughness, toughness indices, and residual strength factors at 28 days**

| ***Specimen*** | $\boldsymbol{T_{\delta_{cr}}}$ N mm | $\boldsymbol{T_{3\delta_{cr}}}$ N mm | $\boldsymbol{T_{5.5\delta_{cr}}}$ N mm | $\boldsymbol{I_5}$ | $\boldsymbol{I_{10}}$ | $\boldsymbol{R_{5-10}}$ |
|---|---|---|---|---|---|---|
| UR (1) | 342 | | | | | |
| UR (2) | 336 | | | | | |
| PA-0.5-1.0% (1) | 492 | 542 | 646 | 2.4 | 4.2 | 36.9 |
| PA-0.5-1.0% (2) | 331 | 537 | 787 | | | |
| PA-0.5-1.5% (1) | 362 | 665 | 860 | 2.9 | 5.4 | 50.7 |
| PA-0.5-1.5% (2) | 418 | 824 | 1126 | | | |
| PA-1.0-1.0% (1) | 637 | 1135 | 1639 | 2.7 | 5.1 | 47.9 |
| PA-1.0-1.0% (2) | 693 | 1079 | 1535 | | | |
| PA-1.0-1.5% (1) | 576 | 944 | 1433 | 2.7 | 5.7 | 59.4 |
| PA-1.0-1.5% (2) | 637 | 1165 | 2203 | | | |
| PA-1.0-1.5% (1) | 418 | 711 | 1211 | 2.9 | 6.1 | 63.0 |
| PA-1.0-1.5% (2) | 399 | 869 | 1362 | | | |
| PA-1.0-1.5% (1) | 712 | 1875 | 3168 | 3.4 | 7.3 | 77.2 |
| PA-1.0-1.5% (2) | 595 | 1333 | 1941 | | | |

As seen in Table 9, the recycled nylon-reinforced mortar exhibits relevant values of residual strength factors $R_{10,5}$. More specifically, we observe remarkable increments of the residual strength factor both on increasing the percentage of fibers (to 1.5 rather than 1.0%) and on increasing the fiber length (to 1.0 and 1.5 in rather than 0.5 in). This circumstance is consistent with what was observed previously in the load–displacement curves (cf. Section 3.2). Marked increases of the material toughness (compared with the unreinforced material) were also observed in the case of a R-PET reinforcement of the same mortar analyzed in the present study [23]. In both cases, the toughness indexes $I_5$ and $I_{10}$ increase with the fiber length (cf. Table 4 in [23] and Table 9 in this document). Toughness increments were also observed by Ozger et al. [25] in the case of concrete reinforcement with nylon fibers obtained from post-consumer textile carpet waste.

## 4 CONCLUSIONS

First, we analyzed the effectiveness of recycled nylon fibers obtained from waste fishing nets as tensile reinforcement for mortars. We then considered the different weight fractions and aspect ratios of these fibers, establishing comparisons with the behavior of unreinforced mortar and with similar investigations available in the literature.

The outcomes of the experimental study indicated that R-Nylon fibers:

- can be safely used as reinforcement in cement materials;
- present adequate alkali resistance of recycled nylon fibers according to currently recognized standards;
- may significantly improve the tensile strength (up to +35%) and fracture properties of cement mortars;
- systematically cause the transformation of a brittle failure mode (unreinforced mortar) to a more ductile failure mode (fiber reinforced mortar).

Our results show that a higher percentage of fibers (1.5% rather than 1.0%) causes a less noticeable drop in the load after the peak value, while higher fiber aspect ratios give the reinforced mortar a hardening-type post-peak behavior. As a consequence, we observed remarkable increments of the toughness indices and of the residual strength factors both when increasing the percentage of fibers (from 1.0% to 1.5%) and when increasing the fiber length (from 0.5 in to 1.0 in and 1.5 in).

On comparing the outcomes of the present study with results given in existing literature [10],[23], [25], we are led to conclude that:

- The toughness and ductility properties of mortars and concretes significantly benefit from the addition of recycled reinforcing fibers to the mix-design;
- R-Nylon fibers are also beneficial in terms of first-crack strength, as opposed to the R-PET fibers analyzed in [23];
- R-Nylon fibers used in this study appear to be much more effective in improving the mechanical qualities of cement products compared with recycled nylon fibers obtained from post-consumer textile carpet waste [25] and recycled PET fibers [23];
- Greater improvements in first-crack strength (up to +149%) and compressive strength (up to +75%) are observed by Habib et al [10] on mortar specimens reinforced with industrial nylon fibers.

It is worth noting the environmental benefits related to the recycling of waste fishing nets,

especially considering that the reinforcement technique examined in the present work does not require any energy consuming processes such as material re-polymerization or extrusion. The nets simply need to be collected, washed, and properly cut to obtain reinforcing fibers, resulting in a substantial reduction in costs and energy consumption.

Proposed expansions of the present research will focus on solving the problem of cutting fishing nets from an industrial perspective. We will also look towards considering a wider range of fiber fractions, as well as other types of binders. The experimental results will provide a basis for the multiscale mechanical modeling of cement materials incorporating R-Nylon reinforcements.

## 5 ACKNOWLEDGMENTS

Support for this work was received from the Italian Ministry of Foreign Affairs, Grant No. 00173/2014, Italy-USA Scientific and Technological Cooperation 2014-2015 (`Lavoro realizzato con il contributo del Ministero degli Affari Esteri, Direzione Generale per la Promozione del Sistema Paese'). The authors would like to thank Omega Plastic srl (Battipaglia, Salerno) for the essential information provided about the employed waste fishing nets, and Carmen Polzone (Geo-Consult srl – Manocalzati, Avellino) for many useful discussions and suggestions about the chemical aspects of this work. Special thanks go to Ada Amendola, University of Salerno, for the valuable support provided during tensile tests on R-nylon fibers.